\documentclass[prl,twocolumn,showpacs,groupedaddress,amsmath]{revtex4}

\usepackage[utf8]{inputenc}
\usepackage[german,american]{babel}
\usepackage{times}
\usepackage[T1]{fontenc}

\usepackage{bm}
\usepackage{amssymb}
\usepackage{graphicx}
\usepackage{dsfont}
\usepackage{epsfig}
\usepackage{color}
\usepackage{ulem}

\newcommand{\ket}[1]{\, | \rm #1 \rangle}

\renewcommand{\vec}[1]{\mathbf{#1}}

\newcommand{\e}{\text{e}}

\newcommand{\be}{\hat b}
\newcommand{\bed}{\hat b^\dagger}
\newcommand{\E}{\hat{\mathcal{E}}}
\newcommand{\Ed}{\hat{\mathcal{E}}^\dagger}

\newcommand{\D}{\hat{\Psi}}
\newcommand{\B}{\hat{\Phi}}
\newcommand{\Dd}{\hat{\Psi}^\dagger}

\newcommand{\pdz}{\frac{\partial}{\partial z}}

\newcommand{\pdt}{\frac{\partial}{\partial t}}

\newcommand{\aB}{a_{\rm b}}

\usepackage{hyperref}

\begin{document}

\title{Wigner Crystallization of Single Photons in Cold Rydberg Ensembles}
\author{Johannes Otterbach}
\email{jotterbach@physics.harvard.edu}
\affiliation{Physics Department, Harvard University, Cambridge 02138, MA, USA}

\author{Matthias Moos, Dominik Muth, Michael Fleischhauer}
\affiliation{Fachbereich Physik und Forschungszentrum OPTIMAS, Technische Universit\"at Kaiserslautern, 67663 Kaiserslautern, Germany}

\date{\today}

\begin{abstract}
The coupling of weak light fields to Rydberg states of atoms under conditions of electromagnetically induced transparency (EIT)
leads to the formation of Rydberg polaritons which are quasi-particles with tunable effective mass and non-local interactions.
Confined to one spatial dimension their low energy physics is that of a moving-frame Luttinger liquid which due to the non-local character of the repulsive interaction can form a Wigner crystal of individual photons. We calculate the Luttinger $K$ parameter using density-matrix renormalization group (DMRG) simulations and find that under typical slow-light conditions kinetic energy contributions
are too strong for crystal formation. However, adiabatically increasing the polariton mass by turning a light pulse into stationary spin excitations allows to generate true crystalline order over a finite length. The dynamics of this process and asymptotic correlations are analyzed in terms of a time-dependent Luttinger theory.
\end{abstract}


\maketitle

The extraordinary properties of Rydberg atoms \cite{Gallagher1994}, such as large dipole-dipole interactions and long life-times, are currently attracting much attention. The interest ranges from quantum information \cite{Saffman2010,Lukin2001, Weimer2010} to  many-body phenomena \cite{Loew2012,Robicheaux2005, Loew2009, Pohl2010, Honer2010, Schwarzkopf2011, Lesanovsky2012, Dudin2012b, Schauss2012, Hoening2013,Grusdt2013}.
So far only few works considered the effect of interactions onto the light fields \cite{Pritchard2010, Olmos2011, Gorshkov2011, Peyronel2012, Bariani2012, Dudin2012}. In recent experiments \cite{Pritchard2010} it was shown that under EIT conditions the Rydberg interaction leads to a non-local, and strongly non-linear behavior of the probe field \cite{Sevincli2011,Petrosyan2011}. This gives rise to, e.g., the formation of a small avoided volume which contains at most one excitation \cite{Gorshkov2011,Peyronel2012}. In the present paper we want to explore the many-body properties on larger length scales. One of the simplest but most dramatic effects resulting from a non-local repulsive interaction is the formation of a Wigner crystal, predicted for electrons in the early days of quantum mechanics \cite{Wigner1934}. We will show that a similar phenomenon can be observed in a dilute 1D gas of photons coupled to Rydberg atoms. The resulting quantum state is highly non-classical and cannot be created in 
conventional Kerr-type point-interacting systems \cite{Chang2008, Angelakis2011}. This has potential applications in photon based quantum communication and information. E.g., the regularity of the photon train can provide high bit rates in quantum repeater protocols and multiplexing.

Under conditions of EIT and small excitation densities, the coupling between photons and Rydberg atoms leads to the formation of light-matter quasi-particles, the so-called dark-state polaritons (DSP) \cite{Fleischhauer2000,Fleischhauer2002}. The DSP follow a non-linear Schr\"odinger-equation with an externally tunable mass and additional strong repulsive and non-local interactions.
We analyze the formation of a quasi-crystalline state of polaritons in 1D using DMRG simulations and time-dependent Luttinger-liquid (LL) theory.
We show that under typical time-independent slow-light conditions the moving-frame ground-state displays density-wave correlations that decay fast in propagation direction due to the small polariton mass.  However, using the external control and making the DSP more massive, i.e., converting them into stationary spin excitations, increases the effect of interactions. Consequently, decelerating a light pulse to a full stop inside a gas of Rydberg atoms \cite{Maxwell2012, Hoffmann2012} can lead to perfect crystalline order over the length of the medium. 
We note that Wigner-crystallization of solitons, representing coherent light pulses, has recently been proposed in \cite{Tercas2013}. In contrast  our approach leads to the crystallization of single excitations, forming a regular train of single photon states upon readout.
\begin{figure}[b]
 \centering
\includegraphics[width=.8\columnwidth]{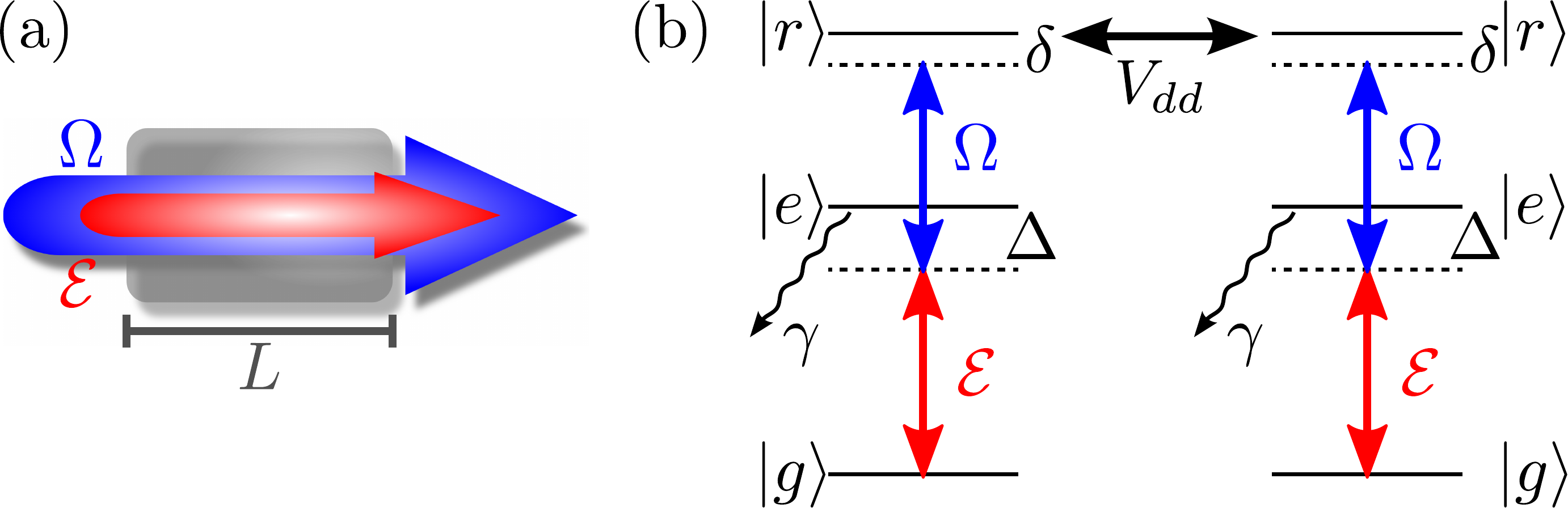}
 \caption{(a) Schematic setup for the creation of dark-state polaritons in a medium on length $L$. (b) Effective atomic linkage pattern for EIT in Rydberg gases. The weak quantized field $\E$ is off-resonantly driving the $\ket{g}-\ket{e}$ transition with a one-photon detuning $\Delta$, whereas the strong control field $\Omega$ is driving the $\ket{e}-\ket{r}$ transition with a final two-photon detuning $\delta$.}
 \label{fig:LadderLinkage}
\end{figure}

To be specific, we consider an ensemble of $N$ atoms with a three-level linkage-pattern [cf. Fig. \ref{fig:LadderLinkage}(b)], composed of a ground-state $\ket{g}$, intermediate state $\ket{e}$ and metastable Rydberg-state $\ket{r}$. The transition $\ket{g}-\ket{e}$ is driven by a quantized probe field $\hat E\,=\,\sqrt{\frac{\hbar\omega_{\rm p}}{2\epsilon_0}} \E(\vec r,t)\e^{-i(\omega_{\rm p} t-\vec{q}_{\rm p}\vec{r})}+\text{H.a.}$, with carrier frequency $\omega_{\rm p}$ and wave-vector $\vec q_{\rm p}$. $\E (\Ed)$ are normalized field amplitudes corresponding to annihilation (creation) of a photon and are slowly varying in space and time. The transition $\ket{e}-\ket{r}$ is coupled via an external control field with Rabi frequency $\Omega$, carrier frequency $\omega_{\rm c}$ and wave-vector $\vec q_{\rm c}$.
We chose the $z$-axis as the common propagation direction of the fields and define the one- and two-photon detunings as $\Delta=\omega_{\rm e}-\omega_{\rm g}-\omega_{\rm p}$, $\delta=\omega_{\rm  r}-\omega_{\rm g}-\omega_{\rm c}-\omega_{\rm p}$, where $\omega_{\rm g,e,r}$ are the energies of the atomic states $(\hbar=1)$. 

In the absence of Rydberg interactions the Hamiltonian can be diagonalized using adiabatic eigen-solutions, the dark- and bright-state polaritons (BSP), which fulfill approximate bosonic commutation relations \cite{Fleischhauer2000, Fleischhauer2002}. Following  \cite{Fleischhauer2008} we define the DSPs as $\D=\cos\theta\E-\sin\theta\hat \Sigma_{\rm gr}$, and BSPs as $\B=\sin\theta\E+\cos\theta\hat \Sigma_{\rm gr}$ where $\tan^2\theta=g^2n/\Omega^2$. Here $\hat \Sigma_{\mu\nu}$ are continuous atomic spin flip operators, $n$ is the atomic density and $g = \wp\sqrt{\omega_p/2\hbar\varepsilon_0}$, with the $\ket{g}-\ket{e}$ dipole moment $\wp$. The DSP propagates lossless with group velocity 
$v_{\rm g}=c\, \cos^2\theta$, while the BSP has a velocity $c\, \sin^2\theta$ and is subject to losses with rate $(g^2 n+\Omega^2)/\Gamma$, where $\Gamma=\gamma + i\Delta$, with $2\gamma$ being the spontaneous decay rate of $|{\rm e}\rangle$.
Near single-photon resonance, $|\Delta|\lesssim \gamma$, and for an optically thick medium, $L\gg L_{\rm abs}$, where $L_{\rm abs}=c\gamma/g^2n$ is the resonant absorption length in absence of EIT and $L$ the medium length, an input bright-polariton will quickly be damped out. In the following we will consider $|\Delta| \gg \gamma$, where absorption is irrelevant. However, for $\cos^2\theta \ll \sin^2\theta\approx 1$ and light pulses of finite length, an input bright-polariton can still be disregarded as it will quickly escape the medium $(c\gg v_{\rm g})$. This allows us to eliminate the BSP and after a short transient the free dynamics is governed by \cite{supplementary}
\begin{eqnarray}
\!\!\!\!\hat H_0 = \int\!\!\text{d}^3\vec{r}\,\Dd(\vec r)\bigg[\frac{\hat p_z^2}{2m_\parallel}+\frac{\hat{\vec p}_\perp^2}{2m_\perp} -v_{\rm g}\hat p_z +\delta(\vec r)\bigg]\D(\vec r),
\label{eq:Dark_FreeEvolution}
\end{eqnarray}
where $\hat p_z=-i \partial_z$, $\hat{\vec p}_\perp=-i\nabla_\perp$, and $\sin^2\theta\approx 1$ was used. This corresponds to an effective Schr\"odinger equation for particles with tensorial mass and additional drift term, moving in an external potential $\delta(\vec r)$. The drift is determined by the EIT group-velocity $v_{\rm g}$, and the masses are $m_\parallel^{-1}=v_{\rm g}L_\text{abs}\frac{\Delta}{\gamma}$ and $m_\perp^{-1}=\frac{v_{\rm g}}{2 q_{\rm p}}$ \cite{Fleischhauer2008, Zimmer2008}. The above model is valid as long as the BSP amplitude is negligible and \cite{supplementary}
\begin{equation}
 |\delta| \ll \frac{g^2 n}{|\Delta|},\qquad \frac{L_\text{abs}}{L_\text{DSP}} \le \frac{\gamma}{|\Delta|}
\label{eq:cond}
\end{equation}
The first condition describes the regime of perturbative coupling between DSP and BSP \cite{Mewes2002}. The second denotes the region of slow-light dispersion \cite{Fleischhauer2002,Zimmer2008}, where $L_\text{DSP}$ is a characteristic length scale of the DSP. Let us now take into account interactions between the atoms in their Rydberg-state $\ket r$, with van-der Waals interaction potential  $V(\vec r)=C_6/|\vec r|^6$. In the continuum limit and transforming to polaritons, we find to lowest order in $\cos\theta$
\begin{align}
 \hat H_\text{int}=\frac{C_6}{2}\int\text{d}^3\vec r\text{d}^3\vec r'\;\frac{\Dd(\vec r) \Dd(\vec r')\D(\vec r')\D(\vec r)}{a^6+|\vec r-\vec r^\prime|^6},
\label{eq:Polariton_Interaction}
\end{align}
where we introduced a cutoff $a$ to account for a possible regularization at short distances \cite{Gorshkov2011}. However, as we will show later, for strong interactions or heavy particles the results become independent of the cutoff and we are allowed to set $a=0$. The effect of the interaction is equivalent to a two-photon detuning. Consequently, the interaction shift has to be smaller than $g^2n/|\Delta|$ which can be translated into a minimal distance $a_{\rm c} = (C_6 |\Delta|/g^2 n)^{1/6}$ the DSPs have to keep to ensure the validity of the model. As shown in  \cite{Gorshkov2011} for the case of a resonant interaction (i.e., $\Delta =0$) and large optical depth, an incoming coherent light pulse will quickly develop strong anti-bunching with a minimum separation length along the propagation direction corresponding to the EIT blockade radius $\aB=(C_6 \gamma/\Omega^2)^{1/6}\gg a_{\rm c}$.  A similar effect happens for $\Delta \ne 0$ due to the fast escape of the BSP. Since under slow-light conditions, $\
cos\theta \ll 1$, the initial preparation produces DSPs with a mutual distance larger than the critical value $a_{\rm c}$ and a vacuum of BSPs, the system is well described by $\hat H=\hat H_0+\hat H_\text{int}$.

To address the question whether the interaction leads to Wigner-crystallization of polaritons we restrict ourselves to one dimension (1D). This can be achieved, e.g., by using elongated cigar-shaped atomic ensembles with transverse extent smaller than the blockade radius \cite{Peyronel2012}, or atoms in hollow-core fibers \cite{Bajcsy2009,Bajcsy2011} or trapped in the evanescent field of ultra-thin optical fibers \cite{Vetsch2010,Dawkins2011}. The low-energy physics can be described in terms of a Luttinger liquid (LL) \cite{Giamarchi-book}. The LL model allows for an exact treatment also in the case of bosons \cite{Citro2007} with $1/|x|^\alpha$-interactions, as long as $\alpha>1$. Transforming to a frame co-moving with the EIT group-velocity removes the drift term, $\propto v_{\rm g}\partial_z$, in eq. \eqref{eq:Dark_FreeEvolution}. Assuming a fixed excitation density $\rho_0$ and $|\Delta| \gg \gamma$, we follow the standard LL-approach \cite{Giamarchi-book} to construct an effective low-energy Hamiltonian
\begin{align}
 H_\text{LL}\,=\,\frac{1}{2\pi}\int\text{d}x\,\left\{uK\left(\pi\hat\Pi\right)^2+\frac{u}{K}\left(\nabla \hat\phi\right)^2\right\}.\label{eq:LL}
\end{align}
$\hat\Pi$ and $\hat\phi$ are conjugate fields with $\left[\hat\phi(x),\hat\Pi(y)\right]=i\delta(x-y)$. $u$ and $K$ are the sound velocity and the Luttinger parameter, respectively. The $K$-parameter governs the asymptotic behavior of the charge-density-wave correlations (CDW) in the ground-state. E.g., the oscillatory part of the density correlations is given by $ \langle\hat\rho(z)\hat\rho(0) \rangle_{\rm osc}\,\sim\, \rho_0^2 \cos(2\pi\rho_0z)\,z^{-2K}$, with $\hat\rho(z)=\Dd(z)\D(z)$. As first-order correlations decay as $\langle\Dd(z)\D(0)\rangle \sim z^{-1/(2K)}$ the point $K=1/2$ marks the crossover from a regime where superfluid order dominates ($K>1/2$) to a regime with predominant CDW correlations of period $1/\rho_0$ ($K<1/2$). We note that technically spoken, the interaction (\ref{eq:Polariton_Interaction}) is of short-range character and we will not find any slower-than-powerlaw correlations as for, e.g., unscreened Coulomb interactions \cite{Schulz1993}.

We like to point out that one can create true crystalline order by adding a weak periodic lattice potential $\delta(x)=\delta_0 \sin(2\pi x/d)$, which leads to a sine-Gordon Hamiltonian \cite{Giamarchi-book} for commensurate fillings $\rho_0=1/(s d)$, $s\in\mathbb{N}$. This model exhibits a quantum phase transition to a gapped ordered phase for arbitrarily small but finite $\delta_0$, if $K<K_s=2/s^2$ \cite{Giamarchi-book,Buechler2011}. To avoid the necessity of a co-moving lattice potential one then should consider stationary-light polaritons \cite{Zimmer2008,Andre-2002,Bajcsy-2003}.

Although no exact expression for $K$ exists, an approximate closed formula was given in \cite{Dalmonte2010}:
\begin{align}
 K=\frac{1}{\sqrt{1+2 \Theta}},\qquad \Theta=\frac{\pi^3}{180}\rho_0^4 m C_6,
\label{eq:K_interpolation}
\end{align}
where $\Theta$ is a measure for the ratio between interaction and kinetic energy.

\begin{figure}[t]
\centering
\includegraphics[width=.8\columnwidth]{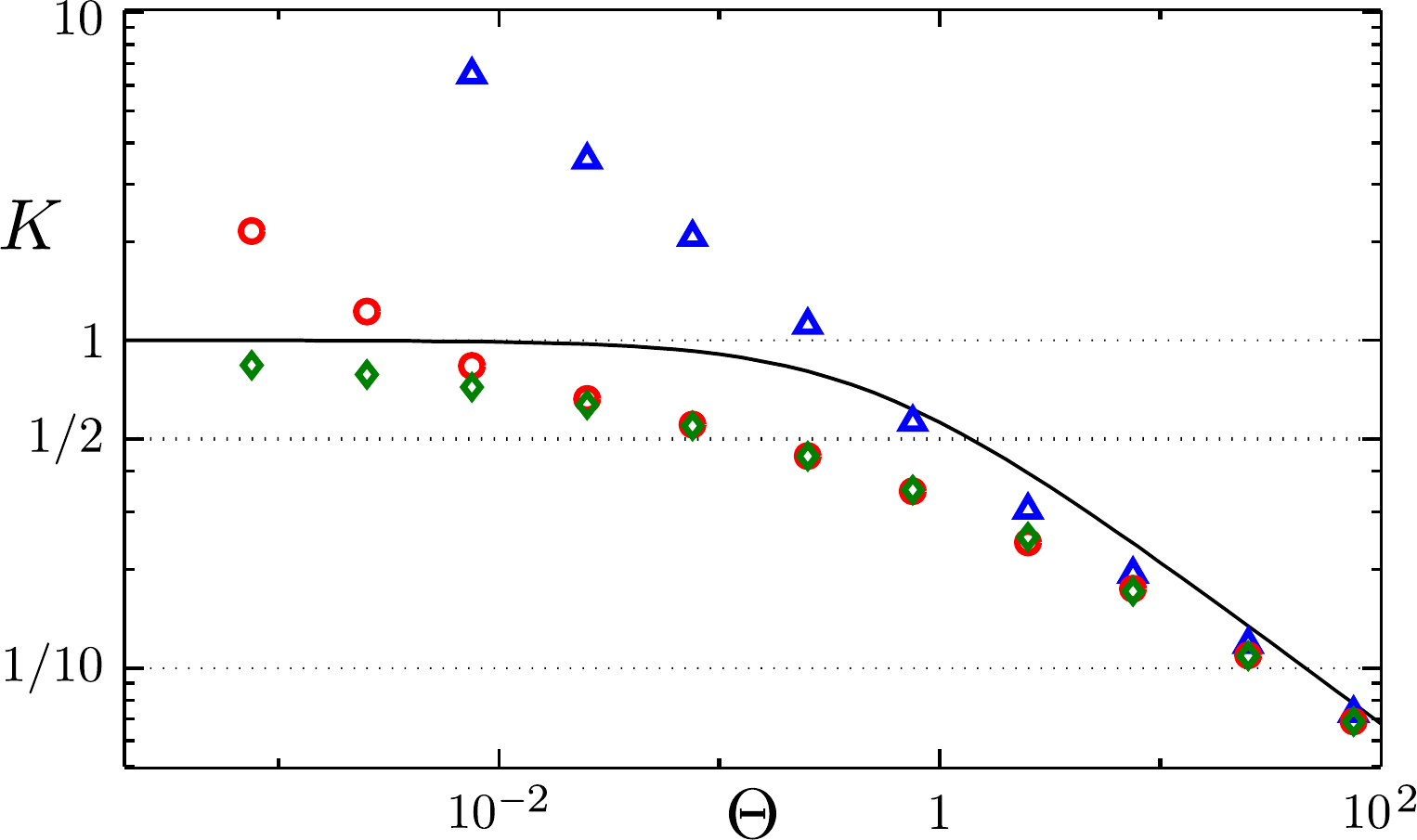}
\caption{Luttinger parameter K as a function of interaction strength. The continuous line shows the analytical approximation \cite{Dalmonte2010} for $C_6/r^6$. Green diamonds show results from DMRG calculations with periodic boundary conditions (BC) for the unscreened optential. Red circles are for $\rho_0 a=1/5$, blue triangles for $\rho_0 a=1/2$. The numerical parameters were $\Delta x=1/10 \rho_0$ with $L=10/\rho_0$ and d=32 for periodic BC. Quantum Monte Carlo results for $C_3/r^3$ can be found in \cite{Citro2007}. }
\label{fig:LLparamsfromDMRG}
\end{figure}

To check this expression we determined $K$ numerically using DMRG \cite{Schollwock2011} and made use of the fact that ${K}/{u} = \pi\rho_0^2\chi$ is determined by the compressibility $\chi^{-1}=\rho_0^2\frac{\partial\mu}{\partial\rho_0} = \rho_0^2L\frac{\partial^2E}{\partial N^2}$ \cite{Giamarchi-book}. Furthermore  $uK = \pi\rho_0/m$, which is true for any Galilean-invariant model \cite{Haldane1981}. We have validated the numerical procedure for the case of the integrable Lieb-Liniger model, where $\chi$ can be calculated exactly as a function of interaction.
Using a proper discretization of the model \cite{Muth2010} leads to the results for $K$ shown in Fig. \ref{fig:LLparamsfromDMRG} for regularized and diverging van-der Waals interactions. As expected for large $\Theta$ (i.e., small $K$) the $K$-parameter becomes independent of the cutoff $a$. Moreover, for $a=0$ eq.(\ref{eq:K_interpolation}) gives the right order of magnitude over the whole range and we will use this expression in the following. As we are interested in the regime of small $K$ we conclude that the actual form of the potential at short distances is irrelevant and only the asymptotic form is important. Hence we can set $a=0$, as long as $\rho_0 a\ll 1$.

\begin{figure}[b]
\centering
\includegraphics[width=.8\columnwidth]{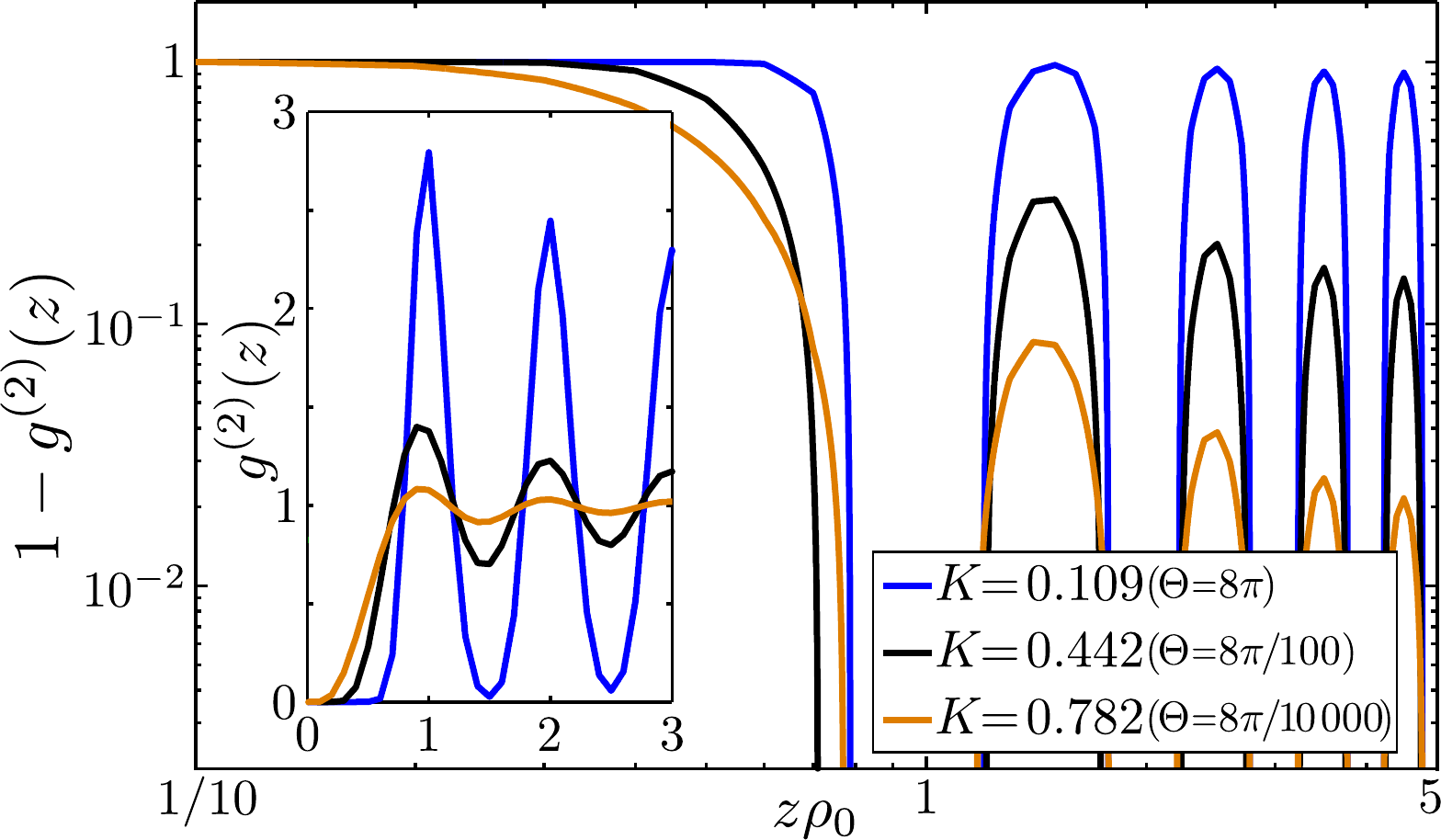}
\caption{Normalized two-particle correlation. {\it main panel:} $1-g^{(2)}(z)$, in double logarithmic scale. {\it inset:} $g^{(2)}(z)$ in linear scale.  Full lines show numerical results for interaction strength increasing from orange over black to blue. The particles are subject to periodic boundary conditions. Note that at $\rho_0 z \gtrsim 5$ finite size effects become noticable. }
\label{fig:CorrelationsDMRG}
\end{figure}

Let us first discuss the time-independent case $\Omega(t) =\Omega=$ const. We concentrate on the ground-state where CDW correlations should be most pronounced. In Fig. \ref{fig:CorrelationsDMRG} we have plotted the normalized two-particle correlation $g^{(2)}(z)$ in the ground state of $\hat H=\hat H_0+\hat H_{\rm int}$ obtained by DMRG corresponding to different values of $K$. The large-$z$ behavior follows the LL expression and one recognizes well pronounced oscillations for $K \le 1/2 $. For small distances the plots show an extended spatial region over which $g^{(2)}(z)$ vanishes, showing that the CDW is a regular array of single-photon Fock states. The correlations around $z=0$ become more suppressed than in the case of free fermions, which is the strongest possible for point interactions \cite{Chang2008, Lieb1963a}. 

We can use eq.\eqref{eq:K_interpolation} to estimate the critical interaction strength required to enter the CDW dominated regime, i.e., $K \leq 1/2$, giving $\Theta_\mathrm{crit}=3/2$.  $\Theta$ is proportional to the effective mass of the polaritons $m\sim v_{\rm g}^{-1} \sim {g^2 n}/{\Omega^2}$ which is different along longitudinal ($m_\parallel$) and transverse directions ($m_\perp$), and can be tuned via the control field $\Omega$. For $m=m_\parallel$ we find
%
\begin{equation}
 \Theta\,=\, \frac{\pi^3}{180}\left(\frac{\gamma}{|\Delta|}\right)^2\left(\rho_0 L_{\rm abs}\right)^4 \frac{c}{v_{\rm g}} \mathrm{OD}_c^6,
\label{eq:Condition_Theta-crit}
\end{equation}
%
where $\mathrm{OD}_c=a_c/L_\text{abs}$ is the optical depth per critical radius. In the crystalline state the characteristic length scale is $L_{\rm DSP} \sim 1/\rho_0$ and thus condition (\ref{eq:cond}) translates into $\rho_0 L_{\rm abs} \lesssim \gamma/|\Delta|$. Using e.g. $\rho_0 L_{\rm abs}=\gamma/|\Delta|=1/100$ and $v_{\rm g}/c =10^{-5}$, we find that the optical depth per critical radius at $\Theta=\Theta_\mathrm{crit}$ has to be OD$_{c}^\parallel\gtrsim 20$. As the mass along the transverse direction is larger, the conditions are more relaxed here and a similar analysis yields OD$_{c}^\perp\gtrsim 5$. Nevertheless, a crystalline structure will be challenging to prepare along both directions as for typical parameters $\mathrm{OD}_c\lesssim 1$. It should be noted, though, that in a finite-size system the CDW might still be observable, as its amplitude can be quite large \cite{Soeffing2009}. 

A closer look at eq. (\ref{eq:Condition_Theta-crit}) suggests a possibility to overcome this challenge using standard light storage techniques \cite{Fleischhauer2000,Phillips2001, Schnorrberger2009}. Let us consider an initial polariton pulse close to the moving-frame ground-state but now with time-dependent control fields. In the absence of interactions, decelerating the DSPs by reducing $v_{\rm g}$ in time preserves their spatial structure and density $\rho_0$ \cite{Fleischhauer2002}. Simultaneously, their effective mass is increased, which suggests an increasing $\Theta$ and hence a decreasing $K$ according to eq.\eqref{eq:K_interpolation} for interacting DSPs. When the pulse is brought to a complete stop, $K(t)$ approaches zero potentially leading to true long-range order. If $v_{\rm g}$ is switched off instantaneously, the initial spatial correlations will be frozen. Thus the switching has to be done smoothly on a time scale $\tau$ long enough for correlations to propagate through the system. The latter 
process is 
determined by the speed of sound $u(t)=\pi\rho_0/(m(t)  K(t))$. For small $K$ we find the scaling $K\sim 1/\sqrt{m} \sim \sqrt{v_{\rm g}}$, i.e., the sound velocity decreases only with the square root of the group velocity, $u(t)\sim 1/\sqrt{m(t)}\sim \sqrt{v_{\rm g}}$, allowing the correlations to propagate through the system before being frozen.

\begin{figure}[t]
\centering
\includegraphics[width=.9\columnwidth]{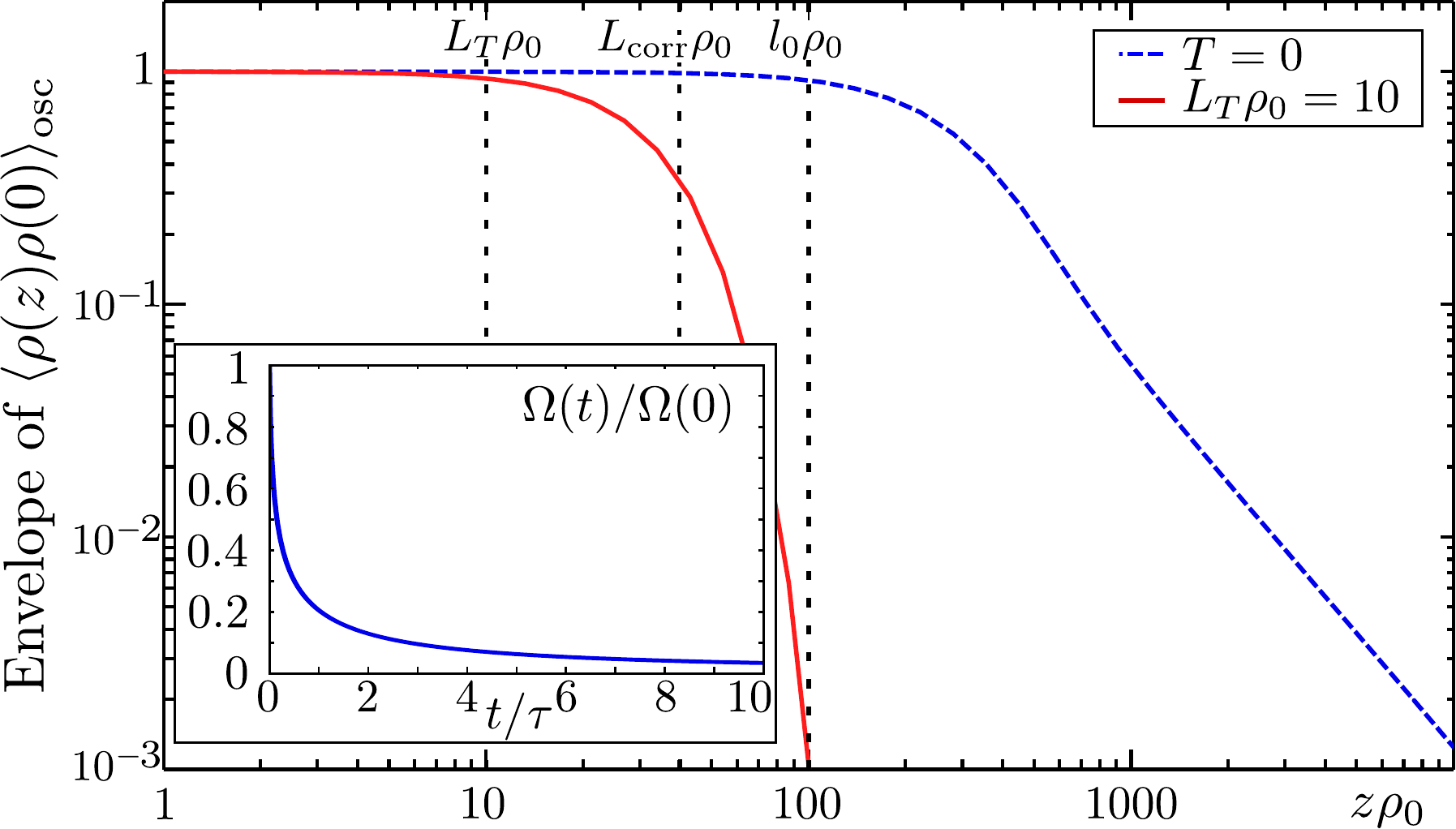}
\caption{{\it main panel:\ }space dependent amplitude of the oscillatory part of the correlation function in the long time limit, $K_0=0.8, K(t)=5\times 10^{-5}$. The dashed blue line shows the spatial decay of density-density correlations for zero Temperature, which shows a crossover from adiabatic to diabatic algebraic decay at $l_0\rho_0=100$ (indicated by rightmost vertical line). The solid red line shows the modified decay for initial Temperature corresponding to a thermal length $L_T\rho_0=10$ (leftmost vertical line) which shows a crossover to exponential decay at length scale $L_\mathrm{corr}\rho_0\approx 40$.\\
{\it inset:\ }$\Omega(t)/\Omega(0)$ for $K_0=0.8$ and $v_\mathrm{gr}(0)/c= 10^{-5}$. }
\label{fig:AsymptoticCorrelations}
\end{figure}

In order to describe the adiabatic switch-off we consider the LL Hamiltonian (\ref{eq:LL}) with time-dependent parameters $K(t)$ and $u(t)$ \cite{Dora2011}. Choosing a special, but generic time-dependence $\Omega(t) = g\sqrt{n}\big/\sqrt{f(t)\,c/v_{\rm g}(0)-1}$, where $f(t) = \e^{x(t)}\,\sinh\left[x(t)\right]\e^{-\mathrm{arcosh}\left(C\right)}/\sqrt{C^2-1}$, with $x(t)\!=\!\mathrm{arcosh}\left(t/\tau+C\right)$ and $C\!=\!(K_0^2+1)/(2 K_0)$, $K_0\!=\!K(t\!=\!0)$ and switch-off time $\tau$, the time-dependent LL model can be solved exactly (see \cite{supplementary} for details).
Due to the finite speed of sound, the final correlations exhibit a ``crossover'' as function of distance from the power-law behavior with adiabatic exponent $K(t)$ to one with the initial exponent $K_0$ at a length-scale $l_0 = \frac{\pi^4}{90} C_6 \rho_0^5 \, \tau\approx\mathrm{OD}_c(\rho_0 a_c)^5 c\tau |\Delta|/\gamma$, as can be seen in Fig. \ref{fig:AsymptoticCorrelations}. Obviously the switch-off time $\tau$  should be maximal. On the other hand, $\tau$ has to be sufficiently small to bring the pulse to a complete stop within the medium, such that $L\leq\int_0^\infty\!\! \mathrm dt\, v_{\rm g}(t)$. Using the above protocol, we find
\begin{equation}
 \frac{l_0}{L} = \frac{2\pi}{K_0}\rho_0L_\text{abs}\frac{|\Delta|}{\gamma}.
\end{equation}
It is interesting to note that this expression does not depend on the interaction strength. This is a consequence of the chosen protocol where the temporal change of $v_{\rm g}(t)$ depends on the interaction strength. Assuming that $K_0$ is close to unity and that $\rho_0 L_{\rm abs} \le \gamma/|\Delta|$, $l_0/L$ can approach unity showing that a crystalline order over the whole medium is possible.

Changing the control-field in time leads to additional couplings between the DSP and BSP \cite{Fleischhauer2002}. The decay rate due to this coupling is given by $\gamma_\theta=\gamma\dot\theta^2/g^2n$. Requiring $\int_0^\tau\!\mathrm{d}t\, \gamma_{\theta}(t)\ll 1$ and using the above protocol we find $c\tau/L_{\text{abs}}\gg 4K_0^2/(K_0^2-1)^2$. For $K_0\approx 0.99$ and $L_\text{abs}\approx 5~\mu\mathrm m$ we have $\tau \gg 0.16\ \mathrm{ns}$ , which is certainly feasible.

So far we have assumed that the initial state for the light storage is the moving-frame ground-state of the LL-Hamiltonian. Let us now discuss the effects of initial excitations. As the system is non-integrable it is reasonable to assume that the  state of the DSPs after the initial preparation is thermal (we set $k_B=1$). In a thermal state all correlations decay exponentially with a correlation length $\sim L_T/K =  \pi\rho_0/(m T K^2)$  \cite{Cazalilla2004}. Since for a non-interacting gas and adiabatic mass changes $T(t)\sim 1/m(t)$ holds, one na\"ively expects that the correlation length increases $\sim 1/K^2(t)$. Evaluating the bosonic correlation functions with a thermal distribution we find a slightly different result (cf. Fig. \ref{fig:AsymptoticCorrelations}). For intermediate length-scales, correlations decay as $\exp(-|z|^2/L_\mathrm{corr}^2)$ as long as $|z|\lesssim L_\mathrm{corr}=2\sqrt{l_0L_T^0}/\pi K_0\times (\ln(K_0/K(t)))^{1/4}$ which crosses over to an exponential decay for larger 
distances \cite{supplementary}. Here $L_T^0$ is the initial thermal correlation length.

To estimate the initial temperature of the DSP we observe that any polariton component with frequency larger than the off-resonant EIT line-width $\Omega^2/|\Delta|$ will escape \cite{supplementary}. Thus a reasonable estimate for an upper temperature limit is $T\lesssim \frac{1}{2} \frac{\Omega^2}{|\Delta|}$ and we obtain the final correlation length at finite temperature as
\begin{align}
 \frac{\sqrt{l_0 L_T^0}}{L}=  \left(\rho_0L_\text{abs}\frac{|\Delta|}{\gamma}\right) \sqrt{\frac{2\pi}{\text{OD}}\frac{|\Delta|}{\gamma}}.
\end{align}
Although the first term on the right side is less than unity, the whole expression can still approach unity and thus even for an initial polariton wave-packet with finite energy an almost perfect Wigner crystal can be created.

An important question is how to observe the crystalline structure of the stored DSPs. This is achieved by turning the stationary excitations back into propagating photons, following standard light retrieval techniques \cite{Fleischhauer2000}. However this must be done by instantaneously switching on the control field as an adiabatic switch-on would just invert the generation process. The resulting regular train of single-photon pulses that leaves the medium can then be detected using standard correlation measurement techniques \cite{Peyronel2012}.

In summary we showed that the combination of EIT with interacting Rydberg gases leads to strongly interacting light-matter particles, termed Rydberg polaritons. We discussed the experimental requirements needed to obtain a quasi-long-range-ordered ground-state corresponding to a moving-frame Wigner crystal of Rydberg excitations in 1D by mapping the problem to a Luttinger liquid. Numerical and analytic results showed that under slow-light conditions the kinetic energy contributions in the longitudinal direction are too large to enter the density-wave dominated regime. Using a time-dependent Luttinger liquid approach we showed, however, that decelerating a light pulse in a gas of Rydberg atoms to a full stop over a sufficiently long deceleration time can create true crystalline order over a substantial fraction of the medium. Turning the Wigner-crystal of spin excitations back into electromagnetic fields by a sudden switch-on of the drive field produces a train of photons with long-range crystalline order.

We thank A.V. Gorshkov, T. Pohl, P. Strack, and M. D. Lukin for fruitful discussions. J.O. acknowledges support by the Harvard Quantum Optics Center. The financial support of the DFG through SFB-TR49 is gratefully acknowledged.


\newpage

\section{Wigner Crystallization of Single Photons in Cold Rydberg Ensemble -- Supplemental Material}

\subsection{Effective Hamiltonian}
In this supplementary we derive the effective Hamiltonian (1) of the main text in one spatial dimension and explicitly establish its conditions of validity. The treatment is based on the formalism developed in \cite{supp-Zimmer2008}. The interaction of probe and control fields with the three-level atoms shown in Fig. 1 of the main text in the absence of Rydberg interactions can be described by the following atom-light coupling Hamiltonian in a rotating frame
\begin{align}
 \hat H = \int &{\rm d}^3\vec{r}\Bigl\{\;\Delta \hat\Sigma_{ee}({\vec r})+\delta\hat\Sigma_{rr}({\vec r}) \\
&+\Omega\hat\Sigma_{re}\e^{i\vec q_c \vec r} +g\sqrt{n} \hat {\cal E}({\vec r})\hat\Sigma_{eg}({\vec r})\e^{i\vec q_p \vec r} + \text{H.a.}\Bigr\},\nonumber
\end{align}
where all quantities are defined as in the main text and  $\hat\Sigma_{\mu\nu}({\vec r}) \equiv \sum_{j\in \Delta V} |\mu\rangle_{jj}\langle \nu|/\sqrt{\Delta N}$ are continuous atomic flip operators defined on a small volume $\Delta V (\vec r)$ centered around position $\vec r$ containing $\Delta N\gg 1$ atoms. Assuming that all atoms are initially prepared in the ground state $|g\rangle$ and considering weak probe fields, i.e. a photon density much less than the atom density, we can treat the light-atom coupling perturbatively. Consequently, in lowest order of the atom-field coupling $g$ we find the Heisenberg-Langevin equations for the atomic operators
\begin{align}
 \pdt\hat\Sigma_{ge} = &-(i\Delta +\gamma)\hat \Sigma_{ge} + \hat F_{ge} \nonumber \\
&+ i g \sqrt{n}\, \hat{\cal E} \e^{i\vec q_p \vec r} + i\Omega^*\, \hat\Sigma_{gr} \e^{-i\vec q_c \vec r},\label{eq:EoM_OpticalPolarization}\\
\pdt\hat\Sigma_{gr} =& -i\delta \hat \Sigma_{gr} + i\Omega\, \hat\Sigma_{ge}\e^{i\vec q_c \vec r}.\label{eq:EoM_SpinPolarization}
\end{align}
Here the $\hat F_{ge}$ is a delta-correlated Langevin noise operator associated with the decay from the intermediate (excited) state $\ket{e}$ which is necessary to preserve commutation relations \cite{supp-Louisell}. One easily verifies that the correlation functions of the Langevin operators are proportional to the population in the excited state $|e\rangle$. In the linear response and for sufficiently small two-photon detuning $\delta$ this population is small and we can safely ignore the noise operators in the following. If need be these operators can be re-introduced by hand using the fluctuation-dissipation theorem. To arrive at a closed description of the atom-field system we also need the equation of motion for the slowly varying probe-field envelope $\hat {\cal E}({\vec r},t)$. Restricting ourselves to a one-dimensional problem the dynamics of the probe field is described by a truncated wave-equation in paraxial approximation
\begin{equation}
 \left[\pdt+c\pdz\right]\E(z,t) = i g\sqrt{n}\, \hat\Sigma_{ge}(z,t)\e^{i\vec q_p \vec r}.\label{eq:EoM_ProbeField}
\end{equation}
Transforming Eqs. (\ref{eq:EoM_OpticalPolarization}-\ref{eq:EoM_ProbeField}) into Fourier space according to $f(z,t) = \int {\rm d}k\, e^{-ikz} f(k,t)$ yields the following matrix equations
\begin{equation}
\frac{d}{dt} \mathbf{X} = -i \sf{H}\mathbf{X}
\end{equation}
where $\mathbf{X}^\top = \{\hat {\cal E},\hat\Sigma_{gr}\e^{i(\vec q_p+\vec q_c) \vec r},\hat\Sigma_{ge}\e^{i\vec q_p \vec r}\}$ and the Hamiltonian matrix reads
\begin{equation}
 \sf{H} = \left[\begin{array}{ccc} -kc & 0 & -g\sqrt{n}\\
                 0 & \delta & -\Omega\\
                -g\sqrt{n} & -\Omega^* & \Delta -i\gamma
                \end{array}\right].\nonumber
\end{equation}
Changing the basis to a description in terms of dark- and bright-polaritons $\mathbf{Y}^\top=\{\hat\Psi,\hat\Phi,\hat\Sigma_{ge}\e^{i\vec q_p \vec r}\}$ via $\hat\Psi = \cos\theta \hat{\cal E} - \sin\theta \hat\Sigma_{gr}\e^{i(\vec q_p+\vec q_c)\vec r} $ and $\hat\Phi = \sin\theta{\cal E}+\cos\theta \hat\Sigma_{gr}\e^{i(\vec q_p+\vec q_c)\vec r} $ yields the equation of motion $\partial_t \mathbf{Y}=-i \sf{H}^\prime\, \mathbf{Y}$ with
\begin{equation}
 \sf{H}^\prime = \left[\begin{array}{ccc}
\delta\sin^2\theta  -kc \cos^2\theta & -\sin\theta\cos\theta(\delta +kc) & 0 \\
-\sin\theta\cos\theta(\delta +kc) & \delta\cos^2\theta  -kc \sin^2\theta & -\Omega_e\\
0 & -\Omega_e &\Delta -i\gamma
                \end{array}\right].\nonumber
\end{equation}
Assuming that the time evolution is slow compared to $|\Delta -i\gamma|$ we can adiabatically eliminate the optical polarization $\hat\Sigma_{ge}$ which yields the coupled equations for bright and dark polaritons
\begin{equation}
 \frac{d}{d t} \left[\begin{array}{c}
                      \hat\Psi \\
		      \hat\Phi
                     \end{array}\right] = -i \sf{H}^{\prime\prime}\, \left[\begin{array}{c}
                      \hat\Psi \\
		      \hat\Phi
                     \end{array}\right]
\end{equation}
with
\begin{equation}
\sf{H}^{\prime\prime} =\left[\begin{array}{cc}
                                                 \delta\sin^2\theta  -kc \cos^2\theta & -\sin\theta\cos\theta(\delta +kc)\\
-\sin\theta\cos\theta(\delta +kc) & \delta\cos^2\theta  -kc \sin^2\theta -\frac{\Omega_e^2}{\Delta -i\gamma}
                                                \end{array}\right]\nonumber.
\end{equation}
For $\Delta > 0$ and under slow-light conditions, i.e. $\sin^2\theta \gg \cos^2\theta$, one recognizes that the off-diagonal
coupling terms are small compared to the difference of the diagonal elements. Under these conditions the dynamics of dark and
bright polaritons approximately separates and one can treat their cross coupling perturbatively. Within this perturbative treatment
the effective equation of motion of the dark polariton $\hat \Psi$ up to second order of the off-diagonal coupling is given by
\begin{eqnarray*}
 \frac{d}{dt}\hat \Psi &=& -i\left(\delta \sin^2\theta - kc\cos^2\theta\right)\hat \Psi \\
&&
- i\frac{\sin^2\theta\cos^2\theta(\delta + kc)^2}{(\delta+kc)(\sin^2\theta - \cos^2\theta) +\frac{\Omega_e^2}{\Delta -i\gamma}}\hat\Psi.
\end{eqnarray*}
To arrive at an even more simplified but more transparent equation we assume $\delta \ge 0$ and require $\delta +kc \ll \Omega_e^2/|\Delta|$ for all relevant values of $k$, which implies in particular
\begin{equation}
 0\le \delta \ll \frac{\Omega_e^2}{|\Delta|}, \label{eq:RestricitonTwoPhotonDetuning}
\end{equation}
where we used $|\Delta|\gg \gamma$. In this limit we find that the dynamics of the DSPs is described by
\begin{eqnarray}
 \frac{d}{dt}\hat \Psi &=& - i\delta\left(1+ \frac{\delta\Delta\cos^2\theta}{\Omega_e^2}\right) \hat\Psi + i k v_{\rm g}\left(1- 2 \frac{\delta \Delta}{\Omega_e^2}\right) \hat \Psi\nonumber\\
&& - i \frac{v_{\rm g} c\Delta}{\Omega_e^2}k^2\hat\Psi,
\end{eqnarray}
where we approximated $\sin^2\theta \approx 1$. The first term on the right hand side describes an energy offset due to a finite two-photon detuning.
The second term accounts for the propagation with group velocity $v_{\rm g}=c\cos^2\theta$. The third term describes the quadratic dispersion
with effective mass $m_\parallel^{-1} = c^2\cos^2\theta |\Delta|/\Omega_e^2 \approx v_{\rm g} L_{\rm abs} |\Delta|/\gamma$.

Condition (\ref{eq:RestricitonTwoPhotonDetuning}) also determines the validity of the interaction Hamiltonian (3) of the main text. As already pointed out in the main text, the Rydberg interactions effectively induce a space-dependent two-photon detuning \cite{supp-Gorshkov2011} which combined with eq. (\ref{eq:RestricitonTwoPhotonDetuning}) leads to the critical minimal distance $a_{\rm c} = (C_\alpha \gamma/\Omega^2_e)^{1/\alpha}$.

We can also interpret the validity of the perturbation theory as a condition of a maximal $k$-value until which
a separation into DSPs with slow-light dispersion and fast moving BSPs  is valid. This condition reads
\begin{align}
 |k c| \ll \frac{\Omega_e^2}{|\Delta|} \label{eq:Restriciton_Kmax}
\end{align}
leading to $|k_{\rm max}| = \Omega_e^2/|\Delta|c$. Plugging this into the zeroth order dispersion relation of the DSP leads to the maximal energy $\omega_{\rm max} = v_{\rm g} k_{\rm max} = \Omega^2/|\Delta|$, which in the end determines the maximal temperature of the DSP gas.

Finally estimating the typical $k$-value of the system via the inverse characteristic length scale, i.e. $k\sim 1/L_{\rm char}$ we can rewrite condition (\ref{eq:Restriciton_Kmax}) and obtain
\begin{align}
 \frac{L_{\rm abs}}{L_{\rm char}}\leq \frac{\gamma}{|\Delta|},
\end{align}
which is just condition (2) of the main text with the appproximation $\sin\theta\approx 1$.

\subsection{Time-dependent Luttinger Liquid theory}
In this section we provide some details of the light storage protocol introduced in the main text and, using time dependent Luttinger liquid (LL) theory, derive an expression for the density-density correlation function.

Decomposition of the LL Hamiltonian (4) from the main text into bosonic momentum modes $\be_p, \bed_p$ \cite{supp-Giamarchi-book} leads to
\begin{equation}
 \hat H=\frac{u(t)}{2}\sum_{p\neq 0}\lvert p\rvert 
\left[w(t)\bed_p\be_p-\frac{g(t)}{2}\left(\bed_p\bed_{-p}\!+\be_{-p}\be{p}\right)\right],
\label{eq:H}
\end{equation} 
where $u(t)=\pi\rho_0/m(t)K(t)$ is the speed of sound and $g(t),\ w(t)$ are given by the LL parameter $K(t)$ as $w(t)=K(t)+1/K(t),\ g(t)=K(t)-1/K(t)$.
The time-dependence of the bosonic operators is given by Heisenberg equations of motion following from
eq.(\ref{eq:H}). To solve these we perform a Bogoliubov transformation relating the time-dependent operators $\bed_p(t),\ \be_p(t)$ to time-independent ones thereby mapping the time-dependence to the coefficients \cite{supp-Dora2011}
\begin{equation}\label{supp:eq:bogoliubov_transformation}
\be_p(t)=u_p(t)\be_p(0)+v_p^*(t)\bed_p(0).
\end{equation}
This yields coupled differential equations for the coefficients 
\begin{equation}\label{supp:eq:de_uv}
 i\partial_t
\begin{pmatrix}u_p(t)\\ v_p(t)\end{pmatrix}
= \frac{\pi\rho_0}{2m(t)}\lvert p\rvert
\begin{pmatrix}w(t)&-g(t)\\ g(t)&-w(t)\end{pmatrix}
\begin{pmatrix}u_p(t)\\ v_p(t)\end{pmatrix}.
\end{equation}
To solve these equations we can diagonalize the coupling matrix on the right hand side. The corresponding transformation matrix is itself time-dependent thus leading 
leading to an off-diagonal coupling $\sim\!\dot K(t)/K(t)$ in the transformed equations
which cannot be neglected. However, if we perform a subsequent diagonalization 
we get an off-diagonal coupling proportional to
\begin{equation}
\frac{\mathrm d}{\mathrm dt}\left(\frac{\dot K(t)}{u(t) K(t)}\right).
\label{eq:nonadiab}
\end{equation}
Since the time-dependence of both, $K(t)$ and $u(t)$, is given by the Polariton mass $m(t)$, we can 
choose the time-dependence such that the off-diagonal coupling vanishes. 
For the choice
\begin{equation}
 m(t)=m_0\frac{e^{2x(t)}-1}{e^{2x(0)}-1},\quad x(t)=\mathrm{acosh}(t/\tau+C),
\end{equation}
with $m_0=m(0)$ and $2C=K(0)+1/K(0)$ expression (\ref{eq:nonadiab}) vanishes.
Inverting the transformations gives an analytic solution for the coefficients $u_p(t),\ v^*_p(t)$. Using \eqref{supp:eq:bogoliubov_transformation} and the fact that the time-independent Hamiltonian for $t\leq 0$ can be diagonalized lets us now compute arbitrary correlation functions. In particular we calculate
\begin{align}\label{supp:eq:G_phi_phi}
 \langle[\phi(z)-\phi(0)]^2\rangle&= \int_0^\infty\mathrm dp\ e^{-\alpha p}\frac{1-\cos 
p z}{p}\nonumber\\ 
&\qquad\times\langle(\bed_p(t)+\be_{-p}(t))(\bed_{-p}(t)+\be_p(t))\rangle\nonumber\\
&=K(t)\bigl[\ln(z/\alpha)+I(z,t)\bigr].
\end{align}
Here $\alpha$ is a high momentum cutoff introduced to treat divergences which we choose as the smallest length scale $\sim\!1/\rho_0$ in our system.
\begin{equation*}
 I(z,t)=\int_0^\infty\!\!\!\mathrm dp\frac{1-\cos
pz}{p}\frac{\cos\xi(t)-1-\sqrt{l_0^2p^2-1}\sin\xi(t)}{1-l_0^2p^2},
\end{equation*}
where $\xi(t)=\sqrt{l_0^2p^2-1}\ln(K(t)/K(0)) $ and can only be evaluated numerically. The expression \eqref{supp:eq:G_phi_phi} allows us to write the oscillatory part of the density-density correlations as follows
\begin{equation}
 \langle\rho(z,t)\rho(0,t)\rangle_\mathrm{osc}\sim\cos(2\pi\rho_0z)\left(\frac{1}{\rho_0z}\right)^{2K(t
)}e^{-K(t)I(z,t)}.
\end{equation}
We see that the algebraic decay with exponent $\sim K(t)$ corresponding to the adiabatic quench gets modified by the exponential $I(z,t)$ which leads to a crossover at length scales $z>l_0$.

\end{document}